\documentclass[12pt]{article}
\usepackage{graphicx}
\usepackage{amsmath}
\usepackage{amssymb}
\usepackage{physics}
\usepackage{authblk}
\usepackage{amsthm}
\usepackage{booktabs}
\usepackage{algorithm}
\usepackage{algpseudocode}

\usepackage{geometry}
\usepackage{setspace}
\usepackage{titlesec}
\usepackage[numbers]{natbib}  
\usepackage[colorlinks=true, linkcolor=black, citecolor=blue]{hyperref}

\titleformat{\section}[block]{\normalfont\Large\bfseries}{\thesection}{1em}{}
\titleformat{\subsection}[block]{\normalfont\large\bfseries}{\thesubsection}{1em}{}
\titleformat{\subsubsection}[block]{\normalfont\normalsize\bfseries}{\thesubsubsection}{1em}{}

\title{\textbf{Consciousness as a Jamming Phase}}

\author[1]{\small Kaichen Ouyang\thanks{Corresponding author: \texttt{oykc@mail.ustc.edu.cn}}}
\affil[1]{\small Department of Mathematics, University of Science and Technology of China, Hefei 230026, China}
\date{}

\begin{document}
\maketitle
\vspace{-1em}  

\begin{abstract}
This paper develops a neural jamming phase diagram that interprets the emergence of consciousness in large language models as a critical phenomenon in high-dimensional disordered systems.By establishing analogies with jamming transitions in granular matter and other complex systems, we identify three fundamental control parameters governing the phase behavior of neural networks: temperature, volume fraction, and stress.The theory provides a unified physical explanation for empirical scaling laws in artificial intelligence, demonstrating how computational cooling, density optimization, and noise reduction collectively drive systems toward a critical jamming surface where generalized intelligence emerges. Remarkably, the same thermodynamic principles that describe conventional jamming transitions appear to underlie the emergence of consciousness in neural networks, evidenced by shared critical signatures including divergent correlation lengths and scaling exponents.Our work explains neural language models' critical scaling through jamming physics, suggesting consciousness is a jamming phase that intrinsically connects knowledge components via long-range correlations.
\end{abstract}


\section{Introduction}
The study of phase transitions and critical phenomena has profoundly shaped our understanding of complex systems across physics, from the spontaneous magnetization of ferromagnets to the superfluidity of liquid helium.Statistical physics provides powerful tools - renormalization group theory, scaling analysis, and universality concepts - that reveal deep organizational principles in seemingly disparate systems\cite{cardy1996scaling}. In recent years, these methods have found surprising applications beyond traditional condensed matter systems, particularly in understanding artificial neural networks\cite{mehta2019high,jiao2024ai}.
A key breakthrough came with the discovery of scaling laws in deep learning, where model performance follows predictable power-law relationships with respect to computational budget, model size, and training data\cite{kaplan2020scaling}.These empirical regularities suggest the existence of underlying phase transitions in the space of neural network parameters. Concurrently, studies of jamming physics have demonstrated how disordered systems - from granular materials to traffic flows - exhibit universal critical behavior when transitioning between fluid-like and solid-like states\cite{o1998jamming,liu2010jamming}.The striking parallels between these domains motivate our central thesis that neural networks undergo a similar critical transition. We propose that generalization - the network's capacity for integrated understanding beyond mere pattern recognition - emerges at a critical jamming point. This transition shares key characteristics with conscious processing in biological systems: (1) global integration of local information, (2) context-dependent flexibility, and (3) the qualitative shift from fragmented to unified representations. The jamming framework thus provides a physical mechanism for how conscious-like generalization emerges in artificial systems through spontaneous symmetry breaking in high-dimensional parameter spaces.

This paper makes three principal contributions:
\begin{itemize}
\item We develop a neural jamming phase diagram with three control parameters ($T_c$, $\phi_c$, $\Sigma_c$) that unifies observed scaling laws in AI with established jamming physics.
\item We demonstrate how computational cooling, density optimization, and noise reduction drive systems toward a critical surface where consciousness emerges.
\item We establish qualitative connections between jamming exponents and neural scaling laws, providing testable predictions for intelligence thresholds.
\end{itemize}

The remainder of this paper is organized as follows: Section 2 reviews related work on scaling laws, long-range correlations, and physical mechanisms of generalization.Section 3 analyzes two paradigmatic jamming systems (bus route model and granular matter) to establish fundamental scaling relationships.Section 4 presents our neural jamming phase diagram and its implications for consciousness.Section 5 concludes with discussion and future directions.
\section{Related Works:}
\paragraph{Scaling Laws:}
Scaling laws represent fundamental power-law relationships that emerge in systems near critical points\cite{domb2000phase}. These universal patterns were first recognized in thermal phase transitions, where physical quantities like correlation length and susceptibility diverge with characteristic exponents as the system approaches critical temperature\cite{brush1967history}. Similar scaling behavior appears in diverse physical systems, from turbulent flows with their distinctive energy spectra\cite{del2004scaling} to granular materials exhibiting singular behavior near jamming transitions\cite{o2003jamming}.The manifestation of scaling laws extends to non-equilibrium systems. Many driven dissipative systems show characteristic power-law dependencies of macroscopic observables on their control parameters\cite{sieberer2013dynamical}. Active matter systems, for instance, demonstrate how collective behavior emerges from simple interaction rules, with measurable quantities following precise scaling relationships\cite{marchetti2013hydrodynamics}.In recent years, analogous scaling laws have been observed in artificial neural networks\cite{bahri2024explaining}. Notably, large-scale language models exhibit consistent power-law relationships between model performance, architecture size, and training data quantity\cite{kaplan2020scaling}. These empirical regularities suggest deep organizational principles shared between complex information processing systems and physical systems displaying critical phenomena.

\paragraph{Long-Range Correlations:} 
Long-range correlations emerge as a fundamental characteristic of systems undergoing phase transitions in statistical physics. Near critical points, local fluctuations can influence distant regions of the system, creating intricate patterns of interdependence that span across scales. This phenomenon appears across nature, from the alignment of magnetic domains in ferromagnetic materials to the coordinated behavior of molecules in liquid crystals\cite{onsager1944crystal,singh2000phase}.The capacity to maintain correlations over large distances extends far beyond equilibrium systems. In biological systems, flocks of birds display remarkable coordination across hundreds of meters through simple interaction rules\cite{toner1995long}. Financial markets exhibit correlations between distant time points through volatility clustering\cite{lux1999scaling}. Even urban traffic patterns show long-range dependencies where local congestion events can propagate influence across entire city networks\cite{o1998jamming}. These diverse examples reveal how complex systems can self-organize to sustain correlations across vast spatial and temporal scales.Modern neural networks have developed sophisticated mechanisms to capture such long-range dependencies.Transformer architectures process relationships between all input elements simultaneously, enabling direct modeling of distant correlations through their attention mechanisms\cite{vaswani2017attention}.Graph neural networks explicitly preserve long-range connectivity through message passing between nodes\cite{gilmer2017neural}.Even convolutional networks, traditionally limited to local receptive fields, achieve global correlation modeling through deep stacking and skip connections\cite{luo2016understanding,he2016deep}. 

\paragraph{Physical Mechanism of Generalization:} 
The generalization mystery of neural networks - why overparameterized models achieve good performance on unseen data - has been progressively decoded through statistical physics frameworks. A key breakthrough came from interpreting grokking as a first-order phase transition, where Rubin et al. demonstrated that neural networks abruptly develop internal representations analogous to mixed-phase formation in physical systems\cite{rubin2023grokking}.This phase transition perspective was further developed by Cohen et al. through field-theoretic methods, showing how renormalization group theory explains learning curves and reveals deep networks' intrinsic bias toward simple polynomial functions.Effective field theories have provided particularly powerful tools for understanding generalization dynamics\cite{cohen2021learning}. Baek and Tegmark's GenEFT framework models latent representations as interacting particles ("repons"), successfully predicting phase transitions between generalization and overfitting regimes\cite{baek2025geneft}.Their approach mirrors the statistical mechanics of many-body systems, where learning rate ratios play the role of temperature-like control parameters. Complementary insights come from evolutionary physics perspectives - Sinitskiy's nonequilibrium optimization theory shows how biological and artificial neural networks share universal scaling properties emerging from selection pressures in high-dimensional parameter spaces\cite{sinitskiy2023making}.These physical approaches collectively reveal that generalization is governed by: (1) phase transitions in representation space, (2) entropy-driven bias toward simple functions, and (3) dynamical criticality in optimization trajectories. The empirical scaling laws discovered by Kaplan et al. ultimately reflect these underlying statistical mechanical principles, where power-law relationships emerge from the self-organization of parameters near marginal stability points\cite{kaplan2020scaling}.

\section{Jamming System and its Scaling Law}
Jamming phenomena ubiquitously emerge in various systems across multiple scales, from granular materials to collective human behaviors. Common examples include sand grains forming stable structures in playgrounds, powders clogging in containers, or traffic congestion during rush hours. While often problematic for industrial processes and daily life, jamming also provides mechanical stability that can be technologically exploited. Although these phenomena have long been observed, their systematic study as a fundamental physical transition only gained prominence in recent decades. In condensed matter physics, jamming refers to the liquid-to-solid transition in disordered systems.This section examines two paradigmatic jamming systems - bus route model\cite{o1998jamming} and granular matter - analyzing their characteristic scaling laws.
\subsection{Bus Route Model and its Dual Model}
\begin{figure}[H]
    \centering
    \includegraphics[width=0.8\textwidth]{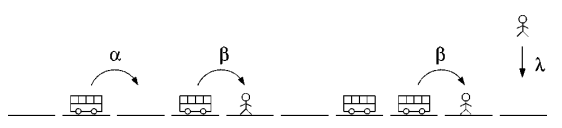}
    \caption{Illustrative Schematic of the BRM Dynamics\cite{o1998jamming}}
    \label{fig:brm_schematic}
\end{figure}

\begin{figure}[H]
    \centering
    \includegraphics[width=0.8\textwidth]{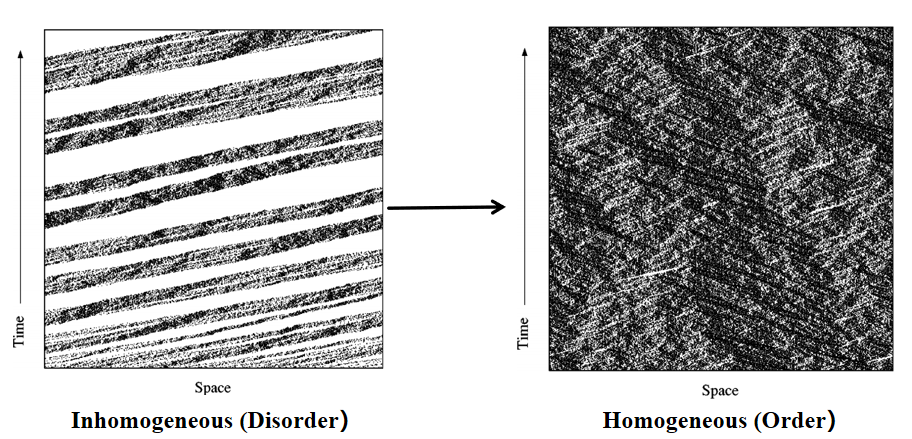}
    \caption{The Disorder-Order Transition in the BRM\cite{o1998jamming}}
    \label{fig:Transition}
\end{figure}

\begin{figure}[H]
    \centering
    \includegraphics[width=0.8\textwidth]{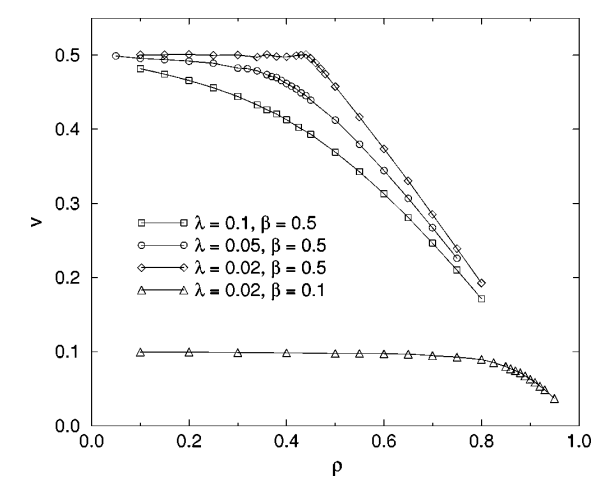}
    \caption{Dependence of velocity on density for different $\lambda$ and $\beta$ values\cite{o1998jamming}}
    \label{fig:vrb}
\end{figure}

\begin{figure}[H]
    \centering
    \includegraphics[width=0.8\textwidth]{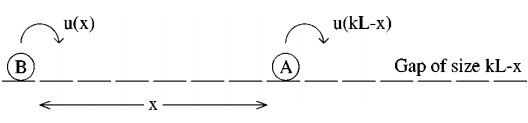}
    \caption{Two-Particle Approximation of the BRM\cite{o1998jamming}}
    \label{fig:2p}
\end{figure}

\paragraph{Bus Route Model:} 
The Bus Route Model (BRM)\cite{o1998jamming} is a one-dimensional driven diffusive system that captures jamming transitions in homogeneous systems. It consists of two interacting components: conserved "buses" (binary variables $\tau_i$) and non-conserved "passengers" (binary variables $f_i$) distributed on a periodic lattice. Buses hop forward with rates $\alpha$ (empty sites) or $\beta<\alpha$ (occupied sites), while passengers appear at empty sites with rate $\lambda$.Figure~\ref{fig:brm_schematic} illustrates the schematic of BRM.
\paragraph{Simulation:}
Numerical simulations reveal that at low bus densities ($\rho < \rho_c = 1-\beta$), the system spontaneously forms bus clusters ("jams") through a coarsening process where smaller jams progressively merge into a single macroscopic cluster, with the leading bus velocity approaching $\beta$ while buses within the jam move at velocity $1-\rho_{\text{jam}}$. Above the critical density ($\rho > \rho_c$), the system maintains a homogeneous bus distribution with collective velocity $1-\rho$, where the transition between these regimes becomes increasingly sharp as the passenger arrival rate $\lambda$ approaches zero, though finite $\lambda$ values show smooth crossover behavior in the velocity-density relationship.Figure~\ref{fig:Transition} demonstrates a hidden disorder-to-order transition in the BRM, while Figure~\ref{fig:vrb} displays the steady-state velocity $v$ as a function of both $\beta$ and $\lambda$, revealing their coupled effects on the system's dynamical behavior.
\paragraph{Two-Particle Approximation:}  
The stability of jams in the BRM can be analyzed through a two-particle approximation examining the gap dynamics between leading buses. Using a mean-field hopping rate $u(x) = \beta + (1-\beta)e^{-\lambda x/\beta}$, the gap size $x$ evolves according to Langevin dynamics in an effective potential $\phi(x)$. For finite $\lambda$, the potential well depth $\Delta \phi \sim \beta(1-\beta)/\lambda$ remains finite in the thermodynamic limit, making jams metastable with exponentially long escape times $\tau \sim L^2 \exp[\beta(1-\beta)/\lambda]$. Only in the $\lambda \to 0$ limit does $\Delta \phi$ diverge, creating truly stable jams. This non-commutativity of limits ($L\to\infty$ vs $\lambda\to 0$) explains why simulations show apparent phase transitions at small but finite $\lambda$, while strict transitions require $\lambda=0$. The approximation breaks down for small $\beta$ due to neglected temporal correlations in bus hopping dynamics.Figure~\ref{fig:2p} illustrates the two-particle approximation scheme in the BRM framework.
\paragraph{Mean-Field Model (MFM) of BRM:}
The mean-field approximation of the BRM replaces the complex many-body dynamics with a solvable hopping particle model where each particle's hopping probability $u(x_i) = \beta + (1-\beta)e^{-\lambda x_i/\beta}$ depends on its forward gap size $x_i$. This MFM maps exactly to a zero-range process with steady-state configuration probability:
\begin{equation}
P(\{x_i\}) = \frac{1}{Z(L,M)}\prod_{i=1}^M q(x_i), \quad \text{where} \quad q(x) = \prod_{y=1}^x \frac{1}{u(y)}
\end{equation}
The partition function $Z(L,M)$ admits a Bose-like interpretation, allowing exact analysis of the jamming transition. While no true phase transition exists for finite $\lambda>0$, the model exhibits critical behavior in the $\lambda\to0$ limit when taking the thermodynamic limit first. The self-consistent determination of mean velocity $v=\langle u\rangle$ provides quantitative agreement with BRM simulations.
\paragraph{Scaling Laws at Small $\lambda$:}  
For finite but small $\lambda$, the MFM exhibits characteristic scaling behavior marking the approach to jamming:
\begin{itemize}
\item Gap distribution $p(x)$ becomes bimodal, developing a secondary peak at $x\sim\xi\sim e^{a/\lambda}$ (with $\xi$ as the characteristic gap size)
\begin{figure}[H]
    \centering
    \includegraphics[width=0.5\textwidth]{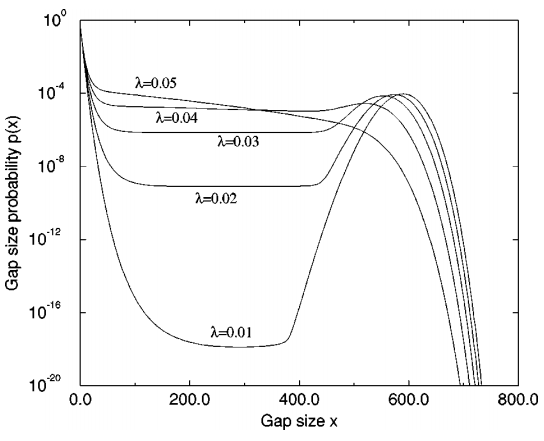}
    \caption{Bimodal behavior in gap size distributions with decreasing $\lambda$\cite{o1998jamming}}
    \label{fig:pxl}
\end{figure}
\begin{figure}[H]
    \centering
    \includegraphics[width=0.5\textwidth]{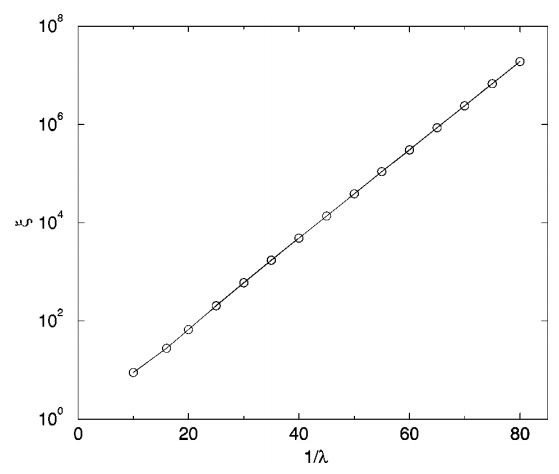}
    \caption{The linear relationship between $\xi$ and $1/\lambda$\cite{o1998jamming}}
    \label{fig:tg}
\end{figure}
\item The probability $P_{ext}$ of extensive gaps follows $P_{ext}M \sim 1$ until system size $L \gg \xi$
\item Velocity curves $v(\rho)$ develop exponentially sharp crossovers near $\rho_c=1-\beta$, with slope $\kappa_{max} \sim e^{b/\lambda}$
\begin{figure}[H]
    \centering
    \includegraphics[width=0.5\textwidth]{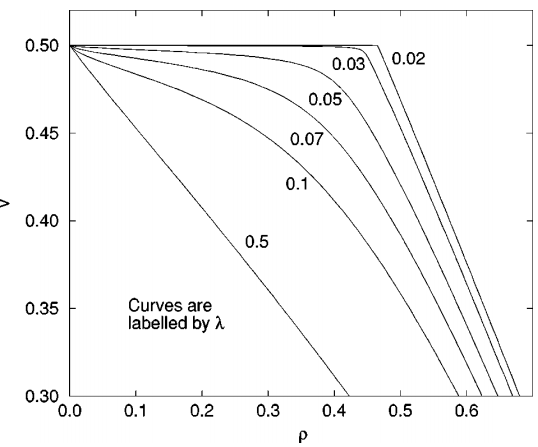}
    \caption{Sharpening of the velocity-density relationship with decreasing $\lambda$\cite{o1998jamming}}
    \label{fig:vr}
\end{figure}
\begin{figure}[H]
    \centering
    \includegraphics[width=0.5\textwidth]{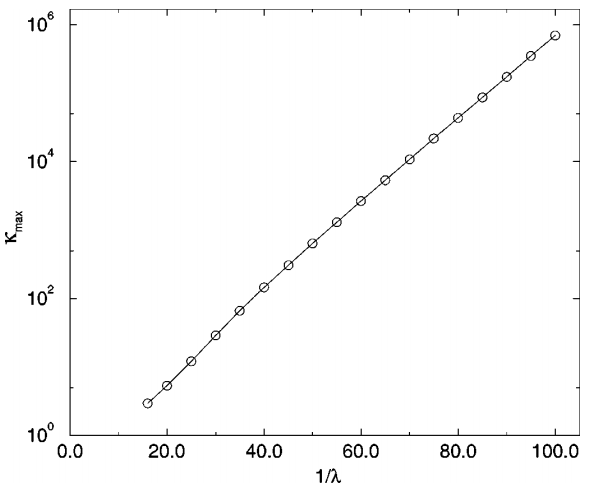}
    \caption{The linear relationship between $\kappa_{max}$ and $1/\lambda$\cite{o1998jamming}}
    \label{fig:kl}
\end{figure}
\end{itemize}
The essential singularity in $\xi$ and crossover sharpness reveals how the $\lambda\to0$ limit restores a true phase transition, while finite $\lambda$ shows pseudocritical behavior with system-size dependent features.
\paragraph{Dual Model:} 
The BRM admits a dual interpretation where holes become mobile particles and buses become empty sites. In this formulation, particles exist in fast ($m_i=1$) or slow ($m_i=\beta$) states, with dynamics: (1) randomly select a particle; (2) fast particles become slow with probability $\lambda$; (3) particles hop left (if space is available) with probability $m_i$; (4) hopping resets the particle to fast state. This describes systems where prolonged stasis reduces mobility, analogous to traffic jams (slow-to-start vehicles) or clogging in narrow pipes (particles sticking to walls). The jamming transition occurs at particle density $1-\rho_c$, manifesting as macroscopic current reduction even for infinitesimal $\lambda$ at high densities. The model captures generic behaviors of driven systems with mobility-dependent interactions.

\subsection{Jamming Transition of Granular Matter}
\begin{figure}[H]
    \centering
    \includegraphics[width=0.5\textwidth]{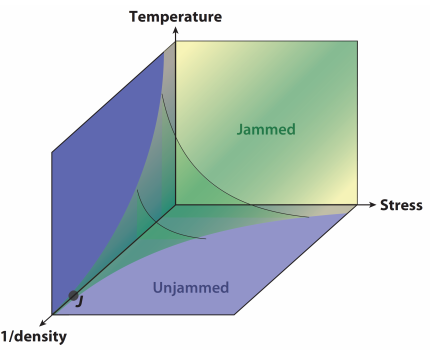}
    \caption{The jamming phase diagram is characterized by three axes: temperature $T$, stress $\Sigma$, and the inverse of packing density $1/\phi$ \cite{liu2010jamming}.}
    \label{fig:jp}
\end{figure}

\paragraph{Jamming Phase Diagram:}
The jamming transition describes how disordered materials change from fluid-like to solid-like behavior. This universal framework connects three fundamental control parameters: temperature ($T$), packing fraction ($\phi$), and applied stress ($\Sigma$). Systems can become jammed (solid-like) through three pathways: (1) lowering $T$ (like supercooled liquids forming glass), (2) increasing $\phi$ (like compressing colloidal particles), or (3) reducing $\Sigma$ (like stopping shear flow in granular materials).Figure~\ref{fig:jp} shows the jamming phase diagram that visually represents these relationships.
In the $T$-$1/\phi$ plane, we observe glass formation when systems avoid crystallization. These systems show characteristic signatures including growing relaxation times and dynamic heterogeneity. While initially hypothesized to be equivalent, glass transitions (dominated by thermal effects) and jamming transitions (dominated by packing effects) are now recognized as distinct phenomena.
The $\Sigma$-$1/\phi$ plane reveals yield stress behavior: when $\Sigma$ exceeds a critical value, the solid begins to flow. Unlike simple Newtonian fluids, jammed systems exhibit complex rheology including shear banding (localized flow regions), shear-rate dependent viscosity (either thinning or thickening), and cooperative particle rearrangements. 
This unified phase diagram has transformed our understanding of disordered systems by revealing deep connections between thermal glasses, athermal granular materials, and colloidal systems - while simultaneously highlighting the rich, non-equilibrium physics that emerges in these diverse systems\cite{liu1998jamming}.
\paragraph{Numerical Simulation Model:}
A typical numerical model for studying jamming physics consists of $N$ soft spherical particles (disks in 2D) with purely repulsive short-range interactions, confined in a periodic box of size $L$ to minimize boundary effects. To prevent crystallization, researchers often use bidisperse systems with equal numbers of large and small particles (typical diameter ratio 1.4:1). Polydisperse systems with Gaussian-distributed diameters can better mimic real-world particle size variations, especially in 2D where monodisperse systems tend to crystallize.
The most common interaction potential is:
\begin{equation}
V(r_{ij}) = \frac{\epsilon}{\alpha}\left(1 - \frac{r_{ij}}{\sigma_{ij}}\right)^\alpha \Theta\left(1 - \frac{r_{ij}}{\sigma_{ij}}\right)
\end{equation}
where $\epsilon$ sets the energy scale, $r_{ij}$ is the interparticle distance, $\sigma_{ij}$ is the sum of particle radii, and $\Theta$ is the Heaviside step function. The exponent $\alpha$ determines the potential shape: $\alpha=2$ gives harmonic springs, while $\alpha=2.5$ yields nonlinear Hertzian contacts.
To generate stable configurations, we start from random particle arrangements and use energy minimization algorithms (like conjugate gradient or FIRE\cite{bitzek2006structural}) to reach mechanically stable states. Below the jamming point ($\phi < \phi_c$), we find unjammed configurations with no particle overlaps. At the jamming transition point $J$, particles just touch with zero energy and pressure. By gradually adjusting particle sizes and repeating minimization, we can create systems at desired pressures or packing fractions above $\phi_c$.

\paragraph{Jamming at Point J:}
The jamming transition point (Point J) represents a special singularity in the jamming phase diagram (Figure 9), governing both the transition itself and material properties across various packing fractions ($\phi$), temperatures ($T$), and shear stresses ($\Sigma$). Determining its location - the critical packing fraction $\phi_c$ - reveals fundamental insights about jamming physics.
For soft sphere systems, statistical analysis shows that while individual configurations jam at different $\phi_c$ values, the probability distribution $P(\phi_c)$ becomes sharply peaked at $\phi^* \approx 0.64$ in the thermodynamic limit. This value remarkably matches experimental random close packing ($\phi_{rcp}$) densities of hard spheres, suggesting a deep connection between maximum entropy states and maximally random jammed configurations.
Two complementary perspectives emerge:
\begin{itemize}
    \item From a geometric viewpoint (using unbiased energy minimization), Point J appears as a well-defined critical point with universal scaling behavior $\xi \sim (\phi - \phi_c)^{-\nu}$, independent of interaction details
    \item From a preparation-history viewpoint, the jamming threshold can vary continuously depending on compression rates or thermal protocols
\end{itemize}
Remarkably, even when $\phi_c$ shifts, systems at Point J maintain isostatic conditions (average contacts $z = 2d$, where $d$ is spatial dimension) and show universal power-law behaviors. This suggests jamming physics extends beyond disordered systems, potentially offering new perspectives on amorphous solids and order-disorder transitions.
\begin{figure}[H]
    \centering
    \includegraphics[width=0.5\textwidth]{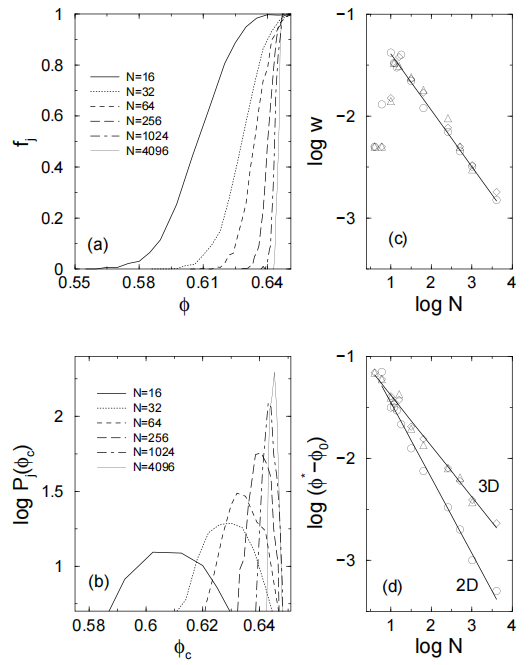}
    \caption{
    (a) Probability $f_j$ of jammed states versus packing fraction $\phi$ for 3D harmonic spheres ($\alpha = 2$). 
    (b) Distribution $P(\phi_c)$ of jamming thresholds from (a). 
    (c) Width $w$ of $P(\phi_c)$ scaling with system size. 
    (d) Finite-size shift of peak position $\phi_0$ relative to $\phi^\star$
    \cite{o2002random}.}
    \label{fig:fp}
\end{figure}

\paragraph{Criticality of Point J:}
The jamming transition exhibits unique critical behavior at point J, where a system of repulsive particles changes from fluid-like to solid-like states. As packing fraction $\phi$ increases beyond $\phi_c$, particles form contacts and the average coordination number $z$ jumps from 0 to $z_c = 2d$ (the isostatic condition), creating a rigid network. This transition shows mixed characteristics:
\begin{itemize}
    \item \textbf{Scaling laws}: Key quantities follow power laws near $\phi_c$:
    \begin{itemize}
        \item Excess contacts $\delta z \equiv z - z_c \sim (\phi - \phi_c)^{1/2}$
        \item Pressure $p \sim (\phi - \phi_c)^{\alpha-1}$
        \item Bulk modulus $B \sim (\phi - \phi_c)^{\alpha-2}$ 
        \item Shear modulus $G \sim (\phi - \phi_c)^{\alpha-3/2}$
    \end{itemize}
    where $\alpha$ depends on interaction potential.
    \item \textbf{Divergent length scales}: Multiple length scales emerge:
    \begin{itemize}
        \item Mechanical length $l^* \sim (\phi - \phi_c)^{-1/2}$ from force balance
        \item Vibrational length $l_T \sim (\phi - \phi_c)^{-1/4}$
        \item Correlation length $\xi \sim (\phi - \phi_c)^{-0.7}$
    \end{itemize}
    \item \textbf{Anomalous features}:
    \begin{itemize}
        \item The ratio $G/B \sim \delta z \to 0$ at point J
        \item No conventional order parameter or structural changes
        \item Combines features of first-order (discontinuous $z$) and second-order (scaling) transitions
    \end{itemize}
\end{itemize}
These properties distinguish jamming from conventional phase transitions and rigidity percolation, despite some similarities. The isostatic condition $z_c = 2d$ plays a central role, governing both the transition point and the emergent mechanical properties of jammed solids.
\begin{figure}[H]
    \centering
    \includegraphics[width=0.4\textwidth]{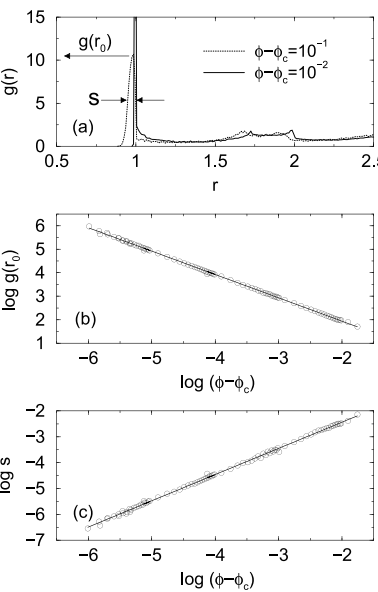}
    \caption{
    (a) Pair correlation function $g(r)$ for a 3D monodisperse harmonic system ($N=1024$), showing definitions of the first peak height $g(r_0)$ and its left-side half-width $s$. 
    (b) Dependence of $g(r_0)$ on $\phi-\phi_c$, with solid line indicating slope $-1$. 
    (c) Dependence of $s$ on $\phi-\phi_c$, with solid line indicating slope $1$
    \cite{o2003jamming}.}
    \label{fig:s1}
\end{figure}
\begin{figure}[H]
    \centering
    \includegraphics[width=0.4\textwidth]{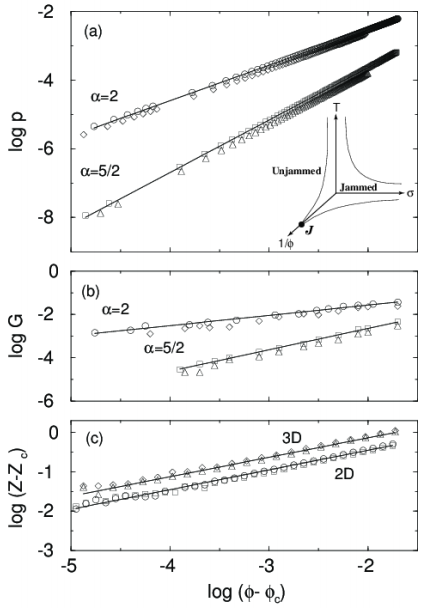}
    \caption{
    (a) Pressure $P$ versus $\phi-\phi_c$ for different dimensions and interaction exponents $\alpha$: circles (squares) show 2D results for $\alpha=2$ ($5/2$), while diamonds (triangles) show 3D results for $\alpha=2$ ($5/2$). System sizes are $N=1024$ (512) for 2D (3D).
    (b) Shear modulus $G$ versus $\phi-\phi_c$.
    (c) Excess contact number $z-z_c$ versus $\phi-\phi_c$
    \cite{o2002random}.}
    \label{fig:s2}
\end{figure}
\section{Neural Jamming Phase Diagram}
\begin{figure}[H]
    \centering
    \includegraphics[width=0.6\textwidth]{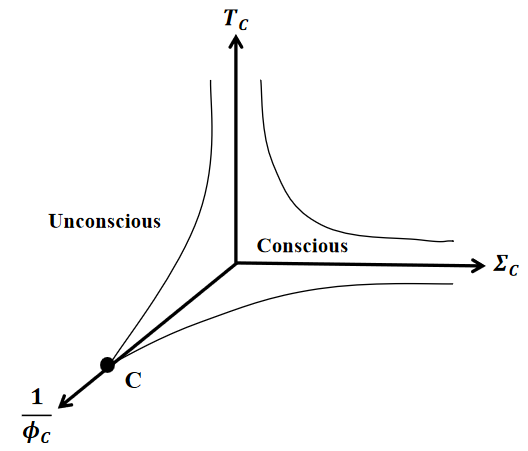}
    \caption{Neural Jamming Phase Diagram}
    \label{fig:njp}
\end{figure}

Inspired by jamming physics, we propose a novel phase diagram for neural language models by treating them as many-particle systems where "atoms" are represented by word embeddings (Fig.~\ref{fig:njp}). This framework phenomenologically identifies three key control parameters:
\paragraph{Effective Temperature ($T_c$)} 
The training process is analogous to thermal annealing in physical systems. The effective temperature $T_c$ is inversely correlated with the total computational budget $C$. Lower $T_c$ corresponds to more extensive training, driving the system from a non-generalizing "fluid" phase into the conscious "jammed" phase through what we term \textit{computational cooling}.
\paragraph{Volume Fraction ($\phi_c$)} 
The system is confined in a fixed high-dimensional hypercube with volume $V_0 \sim \prod_{i=1}^d (u_i - l_i)$, where $d$ is the embedding dimension, and $u_i$, $l_i$ represent the characteristic upper and lower bounds of each dimension in the embedding space. The effective volume occupied by embeddings scales with both model complexity and dataset size as $V_{eff} \propto N_{params} \times N_{data}$. The volume fraction $\phi_c = V_{eff}/V_0$ therefore increases with these two factors, and the jamming transition occurs at point C marking the emergence of generalized intelligence, i.e., consciousness.
\paragraph{Shear Stress ($\Sigma_c$)} 
External perturbations are quantified through an effective shear stress $\Sigma_c$, which captures:
\begin{itemize}
    \item Distributional mismatch: $\Delta p = D_{KL}(p_{train}||p_{real})$
    \item Gradient noise: $\sigma_g^2 = \text{Var}(\nabla_\theta \mathcal{L})$
\end{itemize}
Reducing $\Sigma_c$ facilitates the transition to the jammed phase.
Remarkably, variations along the $T_c$ and $\phi_c$ axes precisely correspond to the empirical scaling laws discovered by Kaplan\cite{kaplan2020scaling}.
The phase diagram suggests that conscious intelligence emerges at the jamming critical point, achievable through:
\begin{itemize}
    \item Computational cooling (increasing $C$ to lower $T_c$)
    \item Density optimization (balancing $N_{params}$ and $N_{data}$ to reach $\rho_c^*$)
    \item Noise reduction (minimizing $\Sigma_c$ via better data/gradient control)
\end{itemize}
The neural jamming phase diagram provides a unified physical explanation for the relationship between model performance (generalization and the emergence of consciousness) and key factors such as model complexity, training data volume, computational budget, and noise in neural language models. This framework enables the analysis of generalization and the conditions for intelligence emergence in neural language models from the perspective of jamming physics.
More importantly, it offers a novel insight into the nature of consciousness—suggesting that consciousness should be understood as a jamming phase. This is because, in the jamming phase, individual content units (word embeddings) develop long-range correlations due to the emergence of generalization, becoming inseparable. The system’s components coalesce into a unified, coherent entity, mirroring the integrated nature of conscious experience.
In this view, consciousness arises when the system reaches a critical jamming state, where global coherence emerges from local interactions—akin to how physical systems exhibit collective behavior at phase transitions. This perspective bridges statistical physics and cognitive science, proposing that intelligence and awareness are natural consequences of a system optimizing toward a high-dimensional, critically jammed state.

\section{Conclusions and Discussion}

We have developed a neural jamming framework that interprets consciousness in language models as an emergent critical phenomenon. By identifying three key control parameters - computational temperature ($T_c$), volume fraction ($\phi_c$), and training stress ($\Sigma_c$) - we establish direct connections between jamming physics and neural scaling laws. The theory explains how increasing computational budgets drive cooling, while optimal model-data balancing achieves critical density for consciousness emergence. Remarkably, the same thermodynamic principles governing granular jamming appear to underlie intelligence emergence in neural networks, evidenced by shared features like isostatic conditions and divergent correlation lengths. This physical perspective not only provides testable predictions for consciousness thresholds but also offers practical guidance for model scaling. The framework suggests that consciousness fundamentally represents a self-organized critical state where local interactions generate global coherence, opening new avenues for understanding intelligence through statistical physics. Future work should focus on quantifying the exact mapping between jamming physics and neural scaling laws, as well as experimentally verifying criticality signatures in large language models.

\bibliographystyle{unsrt}
\bibliography{ref}
\end{document}